\documentclass[prb,amsmath,showpacs]{revtex4}

\begin{document}
%\preprint{}
\title{Conductance and localization in disordered wires: \\ role of evanescent states}
\author{Jean Heinrichs}
\email{J.Heinrichs@ulg.ac.be}
\affiliation{Institut de Physique, B5, Universit\'{e} de Li\`{e}ge, Sart
Tilman, B-4000 Li\`{e}ge, Belgium}

\begin{abstract}
This paper extends an earlier analytical scattering matrix treatment of conductance and 
localization in coupled two- and three Anderson chain systems for weak disorder when evanescent states are present at the 
Fermi level.  Such states exist typically when the interchain coupling exceeds the width of propagating energy bands associated with the various transverse eigenvalues of the coupled tight-binding systems.  We calculate reflection- and 
transmission coefficients in cases where, besides propagating states, one or two evanescent states are available at the 
Fermi level for elastic scattering of electrons by the disordered systems.  We observe important qualitative changes in 
these coefficients and in the related localization lengths due to ineffectiveness of the evanescent modes for transmission 
and reflection in the various scattering channels.  In particular, the localization lengths are generally significantly 
larger than  the values obtained when evanescent modes are absent.  Effects associated with disorder mediated coupling 
between propagating and evanescent modes are shown to be suppressed by quantum interference effects, in lowest order for 
weak disorder.

\end{abstract}

\pacs{72.15.Rn, 73.21.Hb, 73.63.Nm}

\maketitle

\section{Introduction}

In a recent paper \cite{1} (hereafter referred to as I) we developed a scattering matrix analysis of Landauer 
conductance \cite{2} and of localization and reflection in coupled two- and three chain systems for weak disorder, 
using the Anderson tight-binding model.  The full quasi-1D scattering set up consists, as usual, of the system of coupled 
disordered chains of length $L$ connected to electron reservoirs by semi-infinite leads composed of non-disordered
coupled chains attached at both ends.  The leads carry the current which is incident on the disordered sample, together with
the currents which are reflected or transmitted by the sample in modes of properly defined quantum channels \cite{1} at the 
Fermi level.  

The analytical treatment of I was carried out for the case where the states at the Fermi energy in the various channels 
available for scattering of the injected electrons all belong to subbands of propagating states (i.e. the case where all 
channels are propagating).  Thus, in this case all reflected- or transmitted electrons carry current.  On the other hand, 
there may exist Fermi energy domains where the states at the Fermi level corresponding to incident electrons in a particular
propagating subband include evanescent state solutions of the Schr\"{o}dinger equations for some channels (evanescent
channels) besides propagating solutions for others. 

In practice, Fermi energy domains such that propagating states in some channels  coexist at the Fermi level with
evanescent states in other channels are found when the interchain coupling exceeds the width of
the subbands of propagating states in the leads.  Evanescent states do not carry current, as is well-known \cite{3}.  They may 
however influence the current transport indirectly via their coupling, induced by the disorder, to propagating states at the 
Fermi level in other channels.  

The disorder-induced indirect effect of an evanescent channel in the Landauer conductance and in the related localization 
length \cite{4} is quite distinct from the direct effect due to the disappearance of a conducting channel when the Fermi level
is moved accross the edge of its propagating subband into an evanescent domain.  Indeed, this process completely suppresses 
the primary (direct) effect which existed at Fermi energies within the propagating subband of the considered channel, where 
the latter was on the same footing as the other propagating channels coupled to it by the disorder.

The object of this paper is to complement the discussion in I by a detailed analytical study of Landauer conductance and of 
localization lengths, as well as of reflection coefficients in Fermi energy domains where the relevant scattering modes in some
channels are evanescent or non-propagating.  As discussed above evanescent modes generally affect the conductance both
directly, in that their existence implies the absence of corresponding (current carrying) propagating modes, and indirectly via
their coupling to the other propagating modes.

To our knowledge very few studies of the effect of evanescent modes in the conductance or the resistance of disordered wires
have been published, for models  which are different from the Anderson tight-binding model.  Bagwell \cite{5} studied in detail
current transmission amplitudes and electrical conductance as a function of Fermi energy for electrons scattered from a single
$\delta$-function defect (and also numerically, for scattering from a finite range scatterer) in an otherwise non-disordered
multichannel wire.  On the other hand, Cahay {\it et al.} \cite{6} analysed the effect of evanescent states on the resistance
of the two-dimensional  random array of elastic scatterers numerically, using the scattering-matrix formalism of Datta, Cahay
and Mc Lennan \cite{7}.

The scarcity of studies of the effects of evanescent modes  (and the fact that the subject is not mentioned in any of the 
recent reviews or monographs discussing multichannel systems) does not distract from their intrinsic interest.  It suffices 
to recall that the leads in the quantum conductance problem act as electron waveguides which define a basis for the 
scattering matrix of the multichannel disordered region.  This indicates a similarity between the conductance problem and the 
study of optical waveguides where e.g. it is known that both propagating and evanescent wave solutions of the electomagnetic 
wave equation must be superposed in order to correctly describe the fields near sources or obstacles in an otherwise perfect 
waveguide \cite{8}.  On the other hand, in view of the generality of the Anderson model in the context of localization and 
transport in disordered systems, the present study of the role of evanescent modes should be of particular interest.  Also, 
besides their relevance in various experimental situations \cite{1} few channel tight-binding systems are interesting because 
they can be discussed analytically for weak disorder, with the same degree of accuracy as corresponding 1D-systems.

In Sec. II.A we recall the Schr\"{o}dinger tight-binding equations for the two- and three coupled chain systems studied in I.
We briefly discuss the diagonalization of the interchain coupling terms carried out in I, which leads to the description of
the leads in terms of independent channels for scattering of plane wave- and evanescent wave modes.  In the three chain 
case we distinguish between equidistant chains on a planar strip with free boundary conditions and a system of equidistant 
coupled linear chains on a cylindrical surface.  In Sec. II.B we recall the basic formulae for the Landauer two-point 
conductance and for the Lyapounov exponent of the conductance (inverse localization length \cite{4}).  These expressions are 
similar to those used in I except that the summations over channels in which an electron wave may be transmitted (reflected) 
are restricted to the only propagating channels.  In Sec. III we summarize the main points of the determination of the 
transfer- and scattering matrices in these models referring to I \cite{1} for the detailed forms of these matrices 
(with proper adaptations for the case where propagating as well as evanescent channels are present at the Fermi level).  
The final analytic expressions for the averaged coefficients of transmission- and reflection in propagating channels as well 
as expressions for localization lengths are included in Sec. IV.  The discussion of these results together with further
remarks  is given in Sec. V. 

\section{Coupled two- and three chain disordered wires}
\subsection{Tight-binding models in channel bases}

Our $N$-chain model of a wire consists of parallel linear chains of $N_L$ disordered sites each (of spacing $a=1$ and
length $L=N_L a$) connected at both sides to semi-infinite non-disordered $N$-chain leads.

The coupled two-chain Anderson model ($N=2$) is defined by the Schr\"{o}dinger equation in matrix form

\begin{equation}\label{eq1}
\begin{pmatrix}
\varphi^1_{n+1}+\varphi^1_{n-1}\\
\varphi^2_{n+1} +\varphi^2_{n-1}
\end{pmatrix}=
\begin{pmatrix}
E-\varepsilon_{1n} & -h\\
-h & E-\varepsilon_{2n}
\end{pmatrix}
\begin{pmatrix}
\varphi^1_n\\
\varphi^2_n
\end{pmatrix}\quad ,
\end{equation}
where the $\varphi^i_m$ denote the wave-function amplitudes at sites $m$ on the $i$th chain, $h$ is a constant matrix
element for an electron to hop transversally between a site $n$ on chain 1 and the nearest-neighbour site $n$ on chain
2.  The site energies $\varepsilon_{im}$ are random variables associated with the sites $1\leq m\leq N_L$ of the
disordered chain $i$, and $\varepsilon_{im}=0$ on the semi-infinite ideal chains defined by the sites $m>N_L$ and $m<1$,
respectively.  The above energies, including $E$, are measured in units of the constant hopping rate along the
individual chains.

The coupled three chain model ($N=3$) is defined in a similar way by a set of tight-binding equations, whose actual form
depends, however, on interchain boundary conditions.  For free boundary conditions which correspond to arranging the parallel
equidistant chains with nearest-neighbour interchain coupling on a planar strip, the Schr\"{o}dinger equation is

\begin{equation}\label{eq2}
\begin{pmatrix}
\varphi^1_{n+1}+\varphi^1_{n-1}\\
\varphi^2_{n+1} +\varphi^2_{n-1}\\
\varphi^3_{n+1} +\varphi^3_{n-1}
\end{pmatrix}=
\begin{pmatrix}
E-\varepsilon_{1n} & -h & 0\\
-h & E-\varepsilon_{2n} & -h\\
0 & -h & E-\varepsilon_{3n}
\end{pmatrix}
\begin{pmatrix}
\varphi^1_n\\
\varphi^2_n\\
\varphi^3_n
\end{pmatrix}\quad ,
\end{equation}
with the sites in the disordered sections of length $L=N_La$ and in the semi-infinite ideal chain sections labelled in
the same way as in the two-chain case.  On the other hand, in the case of periodic boundary condition which correspond to
equidistant linear chains on a cylindrical surface, the tight-binding equations are

\begin{equation}\label{eq3}
\begin{pmatrix}
\varphi^1_{n+1}+\varphi^1_{n-1}\\
\varphi^2_{n+1} +\varphi^2_{n-1}\\
\varphi^3_{n+1} +\varphi^3_{n-1}
\end{pmatrix}=
\begin{pmatrix}
E-\varepsilon_{1n} & -h & -h\\
-h & E-\varepsilon_{2n} & -h\\
-h & -h & E-\varepsilon_{3n}
\end{pmatrix}
\begin{pmatrix}
\varphi^1_n\\
\varphi^2_n\\
\varphi^3_n
\end{pmatrix}\quad .
\end{equation}

An important aspect of the scattering matrix analysis of conductance in the above systems is the definition of bases
for the scattering matrix which correspond to independent channels at the Fermi energy $E$.  For this purpose one first
transforms the Schr\"{o}dinger equations (\ref{eq1}-\ref{eq3}) to bases in which the interchain hopping terms are
diagonalized.  This transformation, discussed in I, replaces the coupled tight-binding equations describing the leads by
decoupled equations corresponding to independent chain systems or channels.  The transformed Schr\"{o}dinger equations
for the leads in the various systems under consideration are \cite{1}

\begin{equation}\label{eq4}
\begin{pmatrix}
\psi^1_{n+1}+\psi^1_{n-1}\\ \psi^2_{n+1}+\psi^2_{n-1}
\end{pmatrix}=
\begin{pmatrix}
E-h & 0\\
0 & E+h
\end{pmatrix}
\begin{pmatrix}
\psi^1_n\\ \psi^2_n
\end{pmatrix}\quad ,
\end{equation}
for the two-chain system

\begin{equation}\label{eq5}
\begin{pmatrix}
\psi^1_{n+1}+\psi^1_{n-1}\\ \psi^2_{n+1}+\psi^2_{n-1} \\
\psi^3_{n+1}+\psi^3_{n-1}
\end{pmatrix}=
\begin{pmatrix}
e-\sqrt 2 h & 0 & 0\\
0 & E & 0\\
0 & 0 & E+\sqrt 2 h
\end{pmatrix}
\begin{pmatrix}
\psi^1_n\\ \psi^2_n \\ \psi^3_n
\end{pmatrix}\quad ,
\end{equation}
for the three-chain system with the free boundary conditions,

\begin{equation}\label{eq6}
\begin{pmatrix}
\psi^1_{n+1}+\psi^1_{n-1}\\ \psi^2_{n+1}+\psi^2_{n-1} \\
\psi^3_{n+1}+\psi^3_{n-1}
\end{pmatrix}=
\begin{pmatrix}
E-2h & 0 & 0\\
0 & E+h & 0\\
0 & 0 & E+h
\end{pmatrix}
\begin{pmatrix}
\psi^1_n\\ \psi^2_n \\ \psi^3_n
\end{pmatrix},\;
n<1\quad\text{or}\quad n>N_L
\quad ,
\end{equation}
for the three-chain system with periodic boundary conditions.  The transformations $\widehat U^{-1}$ to the new
amplitude vectors

\begin{equation}\label{eq7}
\begin{pmatrix}
\vdots\\\psi_n^i\\\vdots\end{pmatrix}
=\widehat U^{-1}
\begin{pmatrix}
\vdots\\\varphi_n^i\\\vdots\end{pmatrix}\quad ,
\end{equation}
are given  explicitely in I, as are the transformed Schr\"{o}dinger equations for the disordered regions (equations
(8.b-10.b) of I) in which the disordered channels are now coupled by the disorder.

The bases for the definition of transfer matrices and the corresponding scattering matrices in the following section
are provided by the Bloch wave (propagating)- and evanescent mode solutions of the Schr\"{o}dinger eqations
(\ref{eq4}-\ref{eq6}) of the form

\begin{equation}\label{eq8}
\psi^j_n\equiv\psi^{j}_{n,\pm}\sim e^{\pm ik_jn}\quad .
\end{equation}
The propagating solutions for the various channels $j$ correspond to real wavenumbers $k_j$ and exist in energy subbands
defined by

\begin{align}\label{eq9}
2\cos k_1
&=
E-h\quad ,\nonumber\\
2\cos k_2
&=
E+h\quad ,
\end{align}
for the two-channel systems,

\begin{align}\label{eq10}
2\cos k_1
&=
E-\sqrt 2 h\quad ,\nonumber\\
2\cos k_2
&=
E\quad ,\nonumber\\
2\cos k_3
&=
E+\sqrt 2 h\quad ,
\end{align}
for the three-channel system with free boundary conditions, and, finally,

\begin{align}\label{eq11}
2\cos k_1
&=
E-2h\quad ,\nonumber\\
2\cos k_2
&=
2\cos k_3=E+h\quad ,
\end{align}
for the periodic three-channel system.  On the other hand, the evanescent mode solutions for the various channels
correspond to imaginary wavenumbers

\begin{equation}\label{eq12}
k_j=i\kappa_j\quad ,\quad \cos k_j=\cosh \kappa_j\quad ,
\end{equation}
in (\ref{eq8}) and in (\ref{eq9}-\ref{eq11}) i.e. to exponentially growing or decaying solutions,

\begin{equation}\label{eq13}
\psi^j_n\equiv\psi^j_{n,\pm}\sim e^{\mp\kappa_jn}\quad .
\end{equation}
The evanescent modes in a given channel correspond to energies lying outside the energy band of propagating states in
this channel.

\subsection{Conductance and localization}

We describe the conductance of a multichannel disordered wire by the Landauer two-probe formula \cite{2,9}

\begin{equation}\label{eq14}
g=\frac{2e^2}{h}Tr\;(\hat t\hat t^+)\quad .
\end{equation}
Here $\hat t$ denotes the transmission matrix associated with the $M\leq N$ propagating channels of the $N$-channel
system, at the considered Fermi energy,

\begin{equation}\label{eq15}
\hat t=
\begin{pmatrix}
t_{11} & t_{12} & \ldots & t_{1M}\\
t_{21} & \ldots & \ldots & \ldots\\
\ldots & \ldots & \ldots\\
t_{M1} & t_{M2} & \ldots & t_{MM}
\end{pmatrix}\quad ,
\end{equation}
where $t_{ij}$ is the amplitude transmitted in a propagating channel $i$ at the Fermi level at one end of the wire when
there is an incident amplitude in the ( propagating) channel $j$ at the other end.  As in I the localization length is
determined from the rate of exponential decay of the conductance \cite{4,9}, namely

\begin{equation}\label{eq16}
\frac{1}{L_c}=-\lim_{N_L\rightarrow\infty}\frac{1}{N_L}\langle
\ln g\rangle\quad ,
\end{equation}
where averaging over disorder may be used, as usual, because of the self-averaging property of $\ln g$.

The reason for restricting to the propagating channels in the double sum $Tr(\hat t\hat t^+)$ in (\ref{eq14}) is that
evanescent modes at the Fermi level do not contribute to the current.  However, since the conductance of the wire is
defined as the current which flows through it divided by the electro-chemical potential difference between the
reservoirs connected to the disordered region by ideal leads, the evanescent modes may play an indirect role through
their effect on the electrochemical potentials \cite{9,10}.  Such effects have been discussed by Bagwell \cite{5} for
electrons scattered from a single defect in an ideal quasi-1D wire.  Here we consider similar effects for scattering
from coupled chains of random atomic sites.

\section{Scattering Matrix Analysis of Transport}

The analysis of transmission and reflection by the disordered region in I focused on the special case where all the
scattering channels at the Fermi energy are assumed to be propagating.  This situation is encountered typically when
the interchain hopping parameter is less than half the width of the band of propagating states in the various channels
i.e. $|h|<2$.  In the opposite case, $|h|>2$, both evanescent and propagating channels will generally exist at the Fermi
level.

At the level of the general formalism only minor modifications of the analysis of I are required in order to
incorporate the case where evanescent scattering channels may exist.  For discussing this case we shall thus closely
follow the treatment of I and refer to the latter for most of the details.  The first step of the analysis involves
defining transfer matrices for the disordered region of the quasi-1D systems described by (8.b-10.b) of I.  These
transfer matrices are used in a second step for obtaining scattering matrices giving transmission- and reflection
amplitudes which allow us to study the conductance and related properties.

\subsection{Transfer matrices}

Transfer matrices $\tilde Y_n$ for thin slices enclosing only a single site $n$ per channel of the systems described by
(8.b-10.b) of I are defined by rewriting these equations in the form

\begin{equation}\label{eq17}
\begin{pmatrix}
\psi^1_{n+1}\\
\psi^1_{n}\\
\psi^2_{n+1}\\
\psi^2_{n}\\
\vdots
\end{pmatrix}=\tilde Y_n
\begin{pmatrix}
\psi^1_{n}\\
\psi^1_{n-1}\\
\psi^2_{n}\\
\psi^2_{n-1}\\
\vdots
\end{pmatrix}
\quad .
\end{equation}
The matrix $\tilde Y_n\equiv\tilde X_{0n}$ of dimension 4 for the case $N=2$ and the matrices $\tilde Y_n\equiv\tilde
X'_n,\tilde X_n"$ of dimension 6 for the cases $N=3$ with free boundary conditions, respectively, are given  explicitely in
equations (14-16) of I. 

The next step consists in transforming the $\tilde Y_n$ matrices in bases constituted in general by Bloch wave
solutions (equation (\ref{eq8})) of (\ref{eq4}-\ref{eq6}) for some channels and evanescent wave solutions (\ref{eq13})
for the remaining ones at the Fermi level.  In these bases defined by transformation matrices $\widehat W$ in I, the transfer
matrices for the leads denoted, respectively by
$\widehat Y_0\equiv\widehat X_{00},\widehat X'_0,\widehat X_0"$ for the considered cases are diagonal and of the
forms \cite{1}

\begin{equation}\label{eq18}
\widehat Y_0=\text{diag}\;(e^{ik_1},e^{-ik_1},e^{ik_2},e^{-ik_2},\ldots)\quad ,
\end{equation}
where the $k_j$'s, either real or purely imaginary ($k_j=i\kappa_j$), are defined by (\ref{eq9}-\ref{eq11}).  Also, like
in I, real wavenumbers $k_j$ are chosen to be positive, $0\leq k_j\leq\pi$, so that $e^{ik_jn}$ corresponds to a plane
wave moving from left to right.  Similarly a solutions $\psi_n^j=e^{-\kappa_jn}$ ($\kappa_j>0$) is viewed as an
evanescent mode evolving from left to right (and, conversely, $e^{\kappa_jn}$ as a mode evolving from right to left when
discussing scattering from the disordered region.  The full transfer matrices for thin disordered slices (acting on the column
vector $\widehat W^{-1}\left\{\psi^1_{n}, \psi^1_{n-1},\psi^2_{n},\psi^2_{n-1},\ldots\right\}$), in the above mixed Bloch-
and evanescent wave bases, denoted by
$\widehat Y_n\equiv\widehat X_{0n},\widehat X'_n$ and
$\widehat X_n"$ for
$N=2$ and for the two $N=3$ cases, respectively, are given in equations (22) and (23) of I, in terms of tight-binding
parameters defined by (22.a), (24) and (25) of I.  These expressions remain indeed valid for Fermi energies corresponding to
real values of some of the wavenumbers in (\ref{eq9}-\ref{eq11}) and to pure imaginary values for others.

Finally, the transfer matrices for the disordered wires of lengths $L=N_La$ are products of transfer matrices in the
mixed Bloch wave-evanescent wave bases associated with the $N_L$ individual thin slices,

\begin{equation}\label{eq19}
\widehat Y_L=\prod^{N_L}_{n=1}\widehat Y_n\quad .
\end{equation}
As in I, the atomic site energies in (\ref{eq1}-\ref{eq3}) are assumed to be independent gaussian random variables with
zero mean and correlation

\begin{equation}\label{eq20}
\langle\varepsilon_{in}\varepsilon_{jm}\rangle=\varepsilon^2_0\;\delta_{i,j}\;\delta
_{m,n}\quad .
\end{equation}
For weak disorder it is thus sufficient to explicitate (\ref{eq19}) to linear order in the site-energies in order to
study averages to lowest order in the correlation (\ref{eq20}).  The latter implies indeed that different slices in
(\ref{eq19}) are uncorrelated.  Under some notational proviso \cite{11} the final transfer matrices are given by
(30) ($N=2$) and (32) ($N=3$) of I for the real or imaginary wavenumbers $k_1,k_2,k_3$ in
(\ref{eq9}-\ref{eq11}) at an arbitrarily chosen Fermi energy.

\subsection{Scattering matrices}

The scattering of plane waves (reflection and transmission) at and between
the two ends of the random quasi-1D systems is governed by the $S$-matrix,

\begin{equation}\label{eq21}
\widehat S=
\begin{pmatrix}
\hat r^{-+} & \hat t^{--}\\
\hat t^{++} & \hat r^{+-}
\end{pmatrix}\quad ,
\end{equation}
where

\begin{equation}\label{eq22}
\hat t^{\mp\mp}=
\begin{pmatrix}
 t^{\mp\mp}_{11} &  t^{\mp\mp}_{12} & \cdots\\
 t^{\mp\mp}_{21} &  t^{\mp\mp}_{22} & \cdots\\
\vdots & \vdots & \vdots
\end{pmatrix}\quad ,
\end{equation}
and

\begin{equation}\label{eq23}
\hat r^{\pm\mp}=
\begin{pmatrix}
 r^{\pm\mp}_{11} &  r^{\pm\mp}_{12} & \cdots\\
 r^{\pm\mp}_{21} &  r^{\pm\mp}_{22} & \cdots\\
\vdots & \vdots & \vdots
\end{pmatrix}\quad .
\end{equation}
Here $t_{ij}^{++} (t_{ij}^{--})$ and $r_{ij}^{-+}(r_{ij}^{+-})$ denote the
transmitted and reflected amplitudes in channel $i$ (which may be either propagating or evanescent) when there is a unit
flux incident from the left (right) in a current carrying channel $j$.  Left to right- and
right to left directions are labelled + and -, respectively.  The
$S$-matrix expresses outgoing wave amplitudes in terms of ingoing ones on
either side of the quasi-1D disordered wire via the scattering
relations

\begin{equation}\label{eq24}
\begin{pmatrix}
0 \\ 0'
\end{pmatrix}=
\widehat S
\begin{pmatrix}
I \\ I'
\end{pmatrix}\quad .
\end{equation}
Here $I$ and $I'$ (0 and $0'$) denote ingoing (outgoing) amplitudes at the
left and right sides of the disordered region, respectively.  It follows
from current conservation that e.g. for a unit flux which is incident from
the right in channel $i$ one has \cite{9}

\begin{equation}\label{eq25}
\sum^{M}_{j=1}(\mid t^{--}_{ji}\mid^2+\mid r^{-+}_{ji}\mid^2)=1\quad ,
\end{equation}
where the summation is restricted to the $M\leq N$ current carrying channels at the Fermi level.  Likewise, one also
has for propagating channels $j$, 

\begin{equation}
\sum^{M}_{j=1}(\mid t^{++}_{ji}\mid^2+\mid r^{+-}_{ji}\mid^2)=1\quad
\end{equation}

Note that in the case where channels are propagating the current conservation implies that the scattering matrix
(\ref{eq21}) is unitary ($\widehat S\widehat S^+=\widehat 1$).  This is no longer true, of course, in the presence of
evanescent channels. As was shown in I the components of the out- and ingoing waves column vectors in (\ref{eq24}) are given by
the quantities $a^-_{1,0},a^-_{2,0},\ldots,a^-_{N,0},a^+_{1,N_L},a^+_{2,N_L},\ldots,a^+_{N,N_L}$ and 
$a^+_{1,0},a^+_{2,0},\ldots,a^+_{N,0},a^-_{1,N_L},a^-_{2,N_L},\ldots,a^-_{N,N_L}$, respectively, defined
by components of wave transfer column vectors

\begin{equation}\label{eq27}
\{a^+_{1,n-1},a^-_{1,n-1},a^+_{2,n-1},\ldots\}\equiv\widehat W^{-1}\{\psi^1_n,\psi^1_{n-1},\psi^1_n,\ldots\}\quad ,
\end{equation}

and obey the transfer matrix equation for the wires,

\begin{equation}\label{eq28}
\{a^+_{1,N_L},a^-_{1,N_L},a^+_{2,N_L},\ldots\}=\widehat Y_L\{a^+_{1,0},a^-_{1,0},a^+_{2,0},\ldots\}\quad .
\end{equation}

Here the in- and outgoing waves include, in particular, evanescent waves evolving in directions towards- or away from the
disordered wires. The components of the out- and ingoing waves column vectors in (\ref{eq24}) are thus
$a^-_{1,0},a^-_{2,0},\ldots a^-_{N,0},a^+_{1,L},a^+_{2,L},\ldots a^+_{N,L}$ and $a^+_{1,0},a^+_{2,0},\ldots
a^+_{N,0},a^-_{1,L},a^-_{2,L},\ldots a^-_{N,L}$, respectively.  With the so defined vectors of outgoing and incoming
amplitudes, the
$S$-matrix is obtained by rearranging the equation (\ref{eq28}) so as to bring them in the form (\ref{eq24}).  The details of
this somewhat lengthy calculation are explicitated in I.  The explicit forms of the scattering matrices, Eqs (46-47) and (48,
48.a-48.f) of I, for
$N=2$ and
$N=3$, respectively, are expressed in terms of transfer matrix elements which are themselves defined in terms of
general parameters given by (22.a), (24) and (25) of I with wavenumbers, real or pure imaginary, defined by
(\ref{eq9}-\ref{eq11}) above.  These $S$-matrices readily yield the transmission and reflection submatrices in
(\ref{eq21}).

\section{Results}

In the appendix of I we obtained the explicit expressions of transmission- and reflection coefficients,
$|t^{--}_{ij}|^2$ and $|r^{-+}_{ij}|^2$, in terms of tight-binding parameters defining the transfer matrices of the
disordered wires for cases where all channels are propagating at the Fermi level.  In the appendix of the present paper
we discuss the analogous expressions in more general cases where, some of the channels are conducting while the others
are evanescent at the Fermi level.  This requires explicitating the transmission- and reflection amplitude coefficients
($|t^{--}_{ij}|^2$ and $|r^{-+}_{ij}|^2$) for arbitrary wavenumbers, real or imaginary, for which the tight-binding
parameters in (22.a) and (24-25) of I may thus be complex.  In particular, the coefficients $|t^{--}_{ij}|^2$ and
$|r^{-+}_{ij}|^2$ for conducting channels, which are relevant for studying the conductance and related transport
properties, are influenced by the evanescent channels via the disorder mediated coupling between the two types of
channels.

\subsection{Two-channel wires}

The two bands of propagating states in (\ref{eq9}) for real $k_1$, and $k_2$ are non-overlapping for $|h|>2$.  In this
case, to a propagating state in channel 1 corresponds an evanescent state at the same energy in channel 2 and vice
versa.  For the two-channel system the transmission- and reflection coefficients $|t^{--}_{ij}|^2$ and
$|r^{-+}_{ij}|^2$ are given explicitely in (\ref{Appendix1}-\ref{Appendix3}) in the appendix, for the case where
channel 1 is propagating and channel 2 is evanescent.  By expanding these expressions to second order in the
tight-binding parameters (\ref{Appendix3}) and averaging over the disorder, using (\ref{eq20}), we obtain

\begin{equation}\label{eq29}
\langle|t^{--}_{11}|^2\rangle=1-\frac{N_L\varepsilon^2_0}{8\sin^2k_1}\quad ,
\end{equation}

\begin{equation}\label{eq30}
\langle|r^{-+}_{12}|^2\rangle=\frac{N_L\varepsilon^2_0}{8\sin^2k_1}\quad ,
\end{equation}
which obey the current conservation relation (\ref{eq25}) for the conducting channel.

From (\ref{eq29}), (\ref{eq14}) and (\ref{eq16}) it then follows that the weak disorder localization length is

\begin{equation}\label{eq31}
\frac{1}{L_c}=\frac{\varepsilon^2_0}{16\sin^2k_1}\quad .
\end{equation}
The above results will be discussed further in Sec. V.

\subsection{Three-channel wires}

Before discussing coefficients of reflection and transmission of plane waves in various propagating channels when
evanescent modes are present, we first identify the various cases where both types of modes exist in the leads at the
Fermi level, for free-boundary- and periodic systems, respectively.  

{\it Free boundary conditions:} if the subbands of propagating states in (\ref{eq10}) for real $k_1,k_2, k_3$ do note
overlap, that is if

\begin{equation}\label{eq32}
|h|>2\sqrt 2\quad ,
\end{equation}
than to an incident electron in a propagating state of energy $E$ in a particular channel correspond evanescent states
at the same energy in the other two channels.  On the other hand, if the two outer bands 1 and 3 in (\ref{eq10}) do
not overlap while the inner band overlaps with the outer bands i.e. for 

\begin{equation}\label{eq33}
\sqrt 2<|h|<2\sqrt 2\quad ,
\end{equation}
there will be an evanescent mode present at the Fermi level for energies $E$ lying within the overlap regions of the
inner and outer bands.

{\it Periodic boundary conditions}: if the propagating band of channel 1 and the two degenerate bands of channels 2 and
3 in (\ref{eq11}) do note overlap i.e. for

\begin{equation}\label{eq34}
h>\frac{4}{3}\quad ,
\end{equation}
then the waves at the Fermi level in channels 2 and 3 which correspond to an incident wave in the propagating band 1
are evanescent.  Conversely, for incident wave energies lying within the degenerate bands 2 and 3 the modes of the same
energies in channel 1 will be evanescent.  The same is true for $|h|<\frac{4}{3}$ for incident waves of Fermi energy
within the non-overlapping parts of the band of channel 1 and the bands of channels 2, 3, respectively.

In the following we will discuss successively the cases where one or two evanescent modes are present at the Fermi level
for the two types of three-channel systems.  For convenience the expression of transmission and reflection coefficients
in terms of transfer matrix elements of the disordered regions given in (32) of I (under the proviso of  \cite{11}) is
discussed in the appendix and diplayed in detail for the intrachannel transmission coefficients ($|t^{--}_{jj}|^2,
j=1,2,3$ in (\ref{Appendix10}-\ref{Appendix12})).

\subsubsection{Free boundary condition}
a.\;Two evanescent modes

Because of the symmetric arrangement of the propagating bands about $E=0$ we distinguish the cases where the Fermi
level lies in the central $k_2$-band or within one of the outer bands.

For energies within the $k_2$-band, we obtain from (\ref{Appendix11}),

\begin{equation}\label{eq35}
\langle|t^{--}_{22}|^2\rangle=1-\frac{N_L\varepsilon^2_0}{8\sin^2k_2}\quad ,
\end{equation}
and from (48.a), (48.e) and (32) of I,

\begin{equation}\label{eq36}
\langle|r^{-+}_{22}|^2\rangle=\frac{N_L\varepsilon^2_0}{8\sin^2k_2}\quad ,
\end{equation}
using (\ref{eq20}) and (\ref{Appendix8}).  It then follows from (\ref{eq16}) that

\begin{equation}\label{eq37}
\frac{1}{L_c}=\frac{\varepsilon^2_0}{16\sin^2k_2}\quad .
\end{equation}
The current conservation property (\ref{eq25}) for propagating channels is clearly obeyed.

On the other hand, for Fermi energies in the $k_1$-band we get

\begin{equation}\label{eq38}
\langle|t^{--}_{11}|^2\rangle=1-\frac{3N_L\varepsilon^2_0}{32\sin^2k_1}\quad ,
\end{equation}
using (\ref{eq20}) and (\ref{Appendix8}),  and from (48.a), (48.e) and (23) of I,

\begin{equation}\label{eq39}
\langle|r^{-+}_{11}|^2\rangle=\frac{3N_L\varepsilon^2_0}{32\sin^2k_1}\quad ,
\end{equation}
which shows that (\ref{eq25}) is obeyed.  In this case we obtain for the localization length

\begin{equation}\label{eq40}
\frac{1}{L_c}=\frac{3\varepsilon^2_0}{64\sin^2k_1}\quad .
\end{equation}

b.\;One evanescent mode

As a typical case we choose the Fermi level within the overlap region of the $k_1$- and $k_2$-bands, so that the
matching mode in channel 3 is evanescent i.e. $k_3=i\kappa_3$.  In this case we obtain successively from
(\ref{Appendix10}) and (\ref{Appendix11}), using (\ref{eq20}) and (\ref{Appendix8}),

\begin{equation}\label{eq41}
\langle|t^{--}_{11}|^2\rangle=1-\frac{N_L\varepsilon^2_0}{32\sin k_1}
\left(\frac{3}{\sin k_1}+\frac{4}{\sin k_2}\right)\quad ,
\end{equation}

\begin{equation}\label{eq42}
\langle|t^{--}_{22}|^2\rangle=1-\frac{N_L\varepsilon^2_0}{8\sin k_2}
\left(\frac{1}{\sin k_2}+\frac{1}{\sin k_1}\right)\quad ,
\end{equation}
and from (48.a-48.b) and (48.e-48.f) and (32) of I

\begin{equation}\label{eq43}
\langle|t^{--}_{12}|^2\rangle=\langle|t^{--}_{21}|^2\rangle=\frac{N_L\varepsilon^2_0}{16\sin k_1\sin k_2}
\quad ,
\end{equation}
and

\begin{equation}\label{eq44}
\langle|r^{-+}_{11}|^2\rangle=\frac{3N_L\varepsilon^2_0}{32\sin^2 k_1}\quad ,
\end{equation}

\begin{equation}\label{eq45}
\langle|r^{-+}_{22}|^2\rangle=\frac{N_L\varepsilon^2_0}{8\sin^2 k_2}\quad ,
\end{equation}

\begin{equation}\label{eq46}
\langle|r^{-+}_{12}|^2\rangle=\langle|r^{-+}_{21}|^2\rangle=\frac{N_L\varepsilon^2_0}{16\sin k_1\sin k_2}\quad .
\end{equation}

An important check on the correctness of equations (\ref{eq41}-\ref{eq46}) is to note that, again, they verify the
fundamental sum rules (\ref{eq25}) and (25.a) for the currents.  The localization length for weak
disorder obtained from the above results is

\begin{equation}\label{eq47}
\frac{1}{L_c}=\frac{\varepsilon^2_0}{32}\left(\frac{3}{4\sin^2 k_1}+\frac{1}{\sin^2 k_2}
+\frac{1}{\sin k_1\sin k_2}\right)\quad.
\end{equation}

\subsubsection{Periodic boundary conditions}

For periodic boundary conditions, the current conservation properties (\ref{eq25}-25.a) can be obeyed
only if the disorder is restricted to identical realizations in the chains 1 and 2 i.e.

\begin{equation}\label{eq48}
\varepsilon_{1n}=\varepsilon_{2n},\; n=1,2\ldots N_L\quad ,
\end{equation}
while the random site energies on chain 3 remain independent.  This was shown in detail in I, in the case
where all three channels are conducting but remains true as well when evanescent channels are present, as expected.

For energies in the propagating band of channel 1 in cases where there is no overlap with the degenerate propagating
bands 2 and 3 (or for energies in the non-overlapping part of band 1 when this band partly overlaps the bands 2 and 3)
one has evanescent states at the Fermi level in channels 2 and 3.  In this case we obtain, using (\ref{eq20}) and
(\ref{Appendix9}), in (\ref{Appendix10}),

\begin{equation}\label{eq49}
\langle|t^{--}_{11}|^2\rangle=1-\frac{5N_L\varepsilon^2_0}{36\sin^2 k_1}\quad ,
\end{equation}
and in (48.a), (48.e) and (32) of I,

\begin{equation}\label{eq50}
\langle|r^{-+}_{11}|^2\rangle=\frac{5N_L\varepsilon^2_0}{36\sin^2 k_1}\quad .
\end{equation}

By inserting (\ref{eq49}) in (\ref{eq16}) and expanding to linear order in the correlation $\varepsilon^2_0$ we find
for the localization length

\begin{equation}\label{eq51}
\frac{1}{L_c}=\frac{5\varepsilon^2_0}{72\sin^2 k_1}\quad .
\end{equation}

On the other hand, for Fermi energies such that the degenerate channels 2 and 3 are conducting and channel 1 is
evanescent we find, from (\ref{Appendix11}-\ref{Appendix12}),

\begin{equation}\label{eq52}
\langle|t^{--}_{22}|^2\rangle=\langle|t^{--}_{33}|^2\rangle
=1-\frac{N_L\varepsilon^2_0}{4\sin^2 k_2}\quad ,
\end{equation}
and from (48.c), (48.e) and (32) of I,

\begin{equation}\label{eq53}
\langle|t^{--}_{23}|^2\rangle=\langle|t^{--}_{32}|^2\rangle=\frac{N_L\varepsilon^2_0}{18\sin^2 k_2}\quad ,
\end{equation}
using (\ref{eq20}) and (\ref{Appendix9}).  Furthermore, from  (48.a), (48.e) and (32) of I together with (\ref{eq20})
and (\ref{Appendix9}) we get,

\begin{equation}\label{eq54}
\langle|r^{-+}_{22}|^2\rangle=\langle|r^{-+}_{33}|^2\rangle=\frac{5N_L\varepsilon^2_0}{36\sin^2 k_2}\quad ,
\end{equation}

\begin{equation}\label{eq55}
\langle|r^{-+}_{23}|^2\rangle=\langle|r^{-+}_{32}|^2\rangle=\frac{N_L\varepsilon^2_0}{18\sin^2 k_2}\quad .
\end{equation}

The current conservation relation (\ref{eq25}-25.a) are obeyed again and by using the above results in
(\ref{eq16}) we find

\begin{equation}\label{eq56}
\frac{1}{L_c}=\frac{7\varepsilon^2_0}{72\sin^2 k_2}\quad .
\end{equation}

\section{Discussion and concluding remarks}

A simple interpretation of the results of Sect. IV for reflection- and transmission coefficients in two- and three
channel systems in the presence of evanescent channels results from the comparison with the results of I for the case
where all channels are propagating at the Fermi level.  Thus we find that equations (\ref{eq29}) and (\ref{eq30}) for
the two-channel systems ($N=2$) where the channel 2 is evanescent, follow exactly from equation (52) and
(55) of I by suppressing the effect of the channel 2 (or, equivalently, assuming this channel to be absent). 
Similarly, we find that equations (\ref{eq35}) and (\ref{eq36}) for the $N=3$ case with free boundary conditions follow from
(62) and (68) of I by ignoring the effects of channels 1 and 3, which are now assumed to be evanescent.  In the same way
(\ref{eq38}) and (\ref{eq39}) follow from (61) and (67) of I by suppressing the effects of channels 2 and 3.  Finally, in the
case of a single evanescent channel 3 the equations (\ref{eq41}-\ref{eq46}) for the
free boundary model follow from (61-63), (67-68) and (70) of I,
respectively, by ignoring the effect of the channel 3.  Exactly similar conclusions are obtained by comparing the results of
Sec. IV for the $N=3$ case with periodic boundary conditions with the corresponding results of Sec. IV of I.  

As we now discuss these simple properties of the reflection and transmission coefficients in Sec. IV
are partly the consequence of destructive quantum interference, for sufficiently weak disorder of the terms describing the
disorder-mediated coupling between propagating- and evanescent modes at the Fermi level in the corresponding amplitudes of
transmission and reflection.  This is illustrated e.g. in the equations (\ref{Appendix10}-\ref{Appendix12}) for the
random intrachannel transmission coefficients $|t^{--}_{jj}|^2$: the final double sum in each one of these expressions
involves couplings between the propagating channel $j$ and at least one evanescent channel assumed to be present at
the Fermi Level.  Now, since the terms $m=n$ (the only once which survive in the averaging with (\ref{eq20})) in the
sums over coupling terms of propagating and evanescent modes are pure imaginary (see parameters in (\ref{Appendix8})
and (\ref{Appendix9}) for $k_j$ real and $k_i$ imaginary), they add up to zero in the intensity coefficients
(\ref{Appendix10}-\ref{Appendix12}).  In other words, these coupling terms, while existing in the transmission
amplitudes (as they arise from corresponding terms in (\ref{Appendix7})) interfer destructively as a result of their
special phases.

On the other hand, it is clear that the absence of coupling effects between propagating- and evanescent modes in the
results of Sec. IV is specific to our weak disorder approximation.  Indeed corresponding perturbative coupling terms at
higher orders in the parameters (\ref{Appendix8}) and (\ref{Appendix9}) in the transmission- and reflection amplitudes
would have different phases, yielding non vanishing effects in the intensity coefficients.  Similar higher order coupling
effects (of fourth order in the random potential) between a single propagating mode and evanescent modes for a narrow
strip-shaped pure wire with a random boundary ("surface") potential have recently been discussed by Makarov and
Tarasov \cite{12}.

The localization lengths when evanescent modes are included, namely equation (\ref{eq31}) in the two-chain wire,
equations (\ref{eq37}), (\ref{eq40}) and (\ref{eq47}) for the three chain wire with free boundary conditions and,
finally, equations (\ref{eq51}) and (\ref{eq56}) for the three chain case with periodic boundary conditions differ
qualitatively from the results of I (equations (58), (73) and (86) in  \cite{1}, respectively) for the usually
considered case where evanescent states are absent at the Fermi level.  The numerical coefficients in the above
expressions indicate that localization lengths in the presence of evanescent states are generally enhanced by factors
which vary typically between 1 and 2, with respect to corresponding values in I when evanescent states are absent. 
Strongly enhanced localization lengths in the presence of evanescent modes have also been observed in the numerical
calculations of Cahay {\it et al.} \cite{6} for a different model.

The results in Sec. IV for averaged reflection coefficients in the presence of evanescent modes may be used for
deriving mean free paths for elastic scattering of an electron, using the formula \cite{13,14}

\begin{equation}\label{eq57}
\frac{1}{\ell_e}=\frac{1}{MN_L}\sum^M_{i,j=1}\langle|r^{-+}_{ij}|^2\rangle\quad ,
\end{equation}
where $M$ is the number of propagating channels in the quasi-1D conductor.  By inserting the averaged reflection
coefficients for the various quasi-1D systems discussed in Sec. IV, we find in all cases that

\begin{equation}\label{eq58}
\ell_e=\frac{L_c}{2}\quad ,
\end{equation}
where $L_c$ stands for the corresponding localization lengths for weak disorder given by (\ref{eq31}), (\ref{eq37}),
(\ref{eq40}), (\ref{eq47}), (\ref{eq51}) and (\ref{eq52}), respectively.  These $L_c$-values determine
$\frac{1}{\ell_e}$ in the Born approximation for scattering by the disorder.  We note that for systems with a single
propagating channel (\ref{eq58}) coincides formally with the relation between the localization length and the mean free
path in a one-dimensional chain obtained by Thouless \cite{15}.  On the other hand, for systems with two propagating
channels (\ref{eq58}) is analogous to the general formula $L_c=M\ell_c$, of Thouless \cite{16} for a quasi-1D system
with $M$ propagating channels.  The equation (\ref{eq58}) shows that, for weak disorder, the enhancement of the elastic mean
free path due to evanescent states at the Fermi level is proportional to the corresponding enhancement of the localization
length. 

The results of Sec. IV also allow us to express the conditions for the validity of the weak disorder scattering
analysis of transport in quasi-1D systems in a precise physical form.  Clearly our treatment is valid for intrachannel
transmission coefficients close to unity (near-transparency) and sufficiently low reflection- and interchannel
transmission coefficients.  Roughly speaking this requires $L=N_La\leq L_c$ (a condition which can be made more precise for
the various systems and Fermi energy domains by combining the results for the $\langle|t^{--}_{ij}|^2\rangle$ in Sec. IV and
in I with the corresponding expressions for $L_c$), which corresponds to the weak localization or (quasi) metallic regime.  As
is well-known, this regime allows one to find the correct expression for the localization length for weak disorder
($\varepsilon^2_0<<1$).  This has been demonstrated, in particular, for the Anderson model for a one-dimensional chain where
the localization length has been calculated analytically both in the weak- and in the strong localization regime (where all
states are localized on the scale of
$L$ i.e.
$L>L_c$) for weak disorder: the expressions obtained for both cases are identical and coincide with the familiar Thouless
formula, Eq. (51) of I.

In the present paper and in I, we have determined averaged
conductances which allowed us to find localization lengths.  It would be interesting to generalize these analyses to
study conductance fluctuations and the ubiquitous universal conductance fluctuations in the considered coupled Anderson
chain systems.  In particular, since conductance fluctuations involve quartic terms in the tight-binding parameters
(\ref{Appendix8}-\ref{Appendix9}) they would clearly be influenced by the disorder mediated coupling between
evanescent- and propagating modes mentioned earlier in this section.  On the other hand, it would be interesting to
study localization in many-chain tight-binding models of quasi-1D wires.  The exact analytical results obtained in this
paper and in I for two- and three chain systems would be useful limiting cases for future treatments of localization in
many-channel tight-binding wires.

\begin{acknowledgments}
The author would like to thank Professor N. Kumar for encouraging him to investigate evanescent states.
\end{acknowledgments}

\appendix*
\section{Calculation of transmission and reflection coeficients}

As in I we confine ourselves to discussing transmission- and reflection amplitude coefficients $t^{--}_{ij}$ and
$r^{-+}_{ij}$ associated with outgoing waves at the left of the disordered region.

\subsection{Two-channel wires}

In the case where $k_1$ is real and $k_2=i\kappa_2$ is pure imaginary we are interested in finding

\begin{equation}\label{Appendix1}
\mid t_{11}^{--}\mid^2=
\frac{|X_{44}|^2}{|\delta|^2}\quad ,\quad
\mid r_{11}^{-+}\mid^2=
\frac{|X_{21}|^2}{|\delta|^2}
\quad ,
\end{equation}
where $X_{ij}$ denotes the matrix elements of the transfer matrix of the disorder region defined by (30) in I and
$\delta=X_{22}X_{44-}X_{24}X_{42}$.  The generalization of (A.3) of I yields

\begin{align}\label{Appendix2}
|\delta|^2 e^{-2\kappa_2N_L}&=
1+2 \text{Im}\sum^{N_L}_{m=1}a_{2m}
+\sum^{N_L}_{m,n=1}\left[a_{1m}a_{1n}+a_{2m}a^*_{2n}+(b_{m}b_{n}e^{-(ik_1+\kappa_2)(m-n)}+\text{c.c.})\right]\nonumber\\
|X_{44}|^2 e^{-2\kappa_2N_L} &=
1+\sum_{m,n}a_{2m}a_{2n}^*\\
|X_{21}|^2 &=
\sum_{m,n}a_{1m}a_{1n}\cos 2(m-n)k_1\quad , \nonumber
\end{align}
where

\begin{align}\label{Appendix3}
a_{1n}
&=
\frac{\varepsilon_{1n}+\varepsilon_{2n}}{4\sin k_1}\;,\;
a_{2n}\frac{\varepsilon_{1n}+\varepsilon_{2n}}{4i\sinh \kappa_2}\quad , \nonumber
\\
b_n
&=
\frac{\varepsilon_{2n}-\varepsilon_{1n}}{4\sqrt{i\sin k_1\sinh \kappa_2}}\quad .
\end{align}

\subsection{Three-channel wires}

The transmission- and reflection amplitudes of interest defined by the matrices $\widehat S_1$ and $\widehat S_3$ in
(48.a) and (48.e) of I yield

\begin{align}\label{Appendix4}
|t^{--}_{ij}|^2=\frac{|\alpha_{ij}|^2}{\Delta^2}\quad ,
\end{align}
where the $\alpha_{ij}$ are given by quantities $\beta_{k}$ defined in I: $\alpha_{11}=\beta_1,
\alpha_{12}=\beta_4,\alpha_{13}=\beta_7$,\linebreak
$\alpha_{21}=\beta_2,\alpha_{22}=\beta_5,\alpha_{23}=\beta_8,
\alpha_{31}=\beta_3,\alpha_{32}=\beta_6,\alpha_{33}=\beta_9$.

To second order in the site energies the expression (\ref{Appendix4}) reduce to

\begin{align}\label{Appendix5}
\mid r_{11}^{-+}\mid^2
&=
|e^{ik_1N_L}|^2
\mid Y_{21}\mid^2\;,\;
\mid r_{12}^{-+}\mid^2=
|e^{ik_1N_L}|^2
\mid Y_{23}\mid^2\;,\;
\mid r_{13}^{-+}\mid^2=
|e^{ik_1N_L}|^2
\mid Y_{25}\mid^2\quad ,\nonumber \\
\mid r_{21}^{-+}\mid^2
&=
|e^{ik_2N_L}|^2
\mid Y_{41}\mid^2\;,\;
\mid r_{22}^{-+}\mid^2=
|e^{ik_2N_L}|^2
\mid Y_{43}\mid^2\;,\;
\mid r_{23}^{-+}\mid^2=
|e^{ik_2N_L}|^2
\mid Y_{45}\mid^2\quad ,\nonumber \\
\mid r_{31}^{-+}\mid^2
&=
|e^{ik_3N_L}|^2
\mid Y_{61}\mid^2\;,\;
\mid r_{32}^{-+}\mid^2=
|e^{ik_3N_L}|^2
\mid Y_{63}\mid^2\;,\;
\mid r_{33}^{-+}\mid^2=
|e^{ik_3N_L}|^2
\mid Y_{65}\mid^2
\quad .
\end{align}
Here the $Y_{ij}$ are matrix elements of the transfer matrix (32) in I of the disordered region and the $\beta_i$'s and
$\Delta$ are quadratic and cubic forms in these elements defined in (48.e) and (48.f) of I.  The exponential coefficients in
(\ref{Appendix5}) differ from unity for imaginary wavenumbers.

In the three-channel case we have a variety of different domains of Fermi energies where besides one propagating
channel at least, there exist different evanescent channels, both for free boundary conditions and for periodic
boundary conditions.  Therefore we wish to explicitate the expressions (\ref{Appendix4}) and (\ref{Appendix5}) to second
order in forms valid for arbitrary wavenumbers $k_1,k_2,k_3$, real or pure imaginary, which implies that the
tight-binding parameters involved in the transfer matrix (32) of I will be generally complex too.

On the other hand, for weak disorder, we only require the explicit forms of the quantities in
(\ref{Appendix4}-\ref{Appendix5}) to second order in the site energies or, equivalently, to second order in the
tight-binding parameters (24-25) of I.  These explicit forms follow trivially for the reflection coefficients in
(\ref{Appendix5}) and for the interchannel transmission coefficients in (\ref{Appendix4}) since the elements of the
reflection matrix $\hat r^{-+}$ given by (48.a) of I as well as the off-diagonal elements of the transmission matrix
$\hat t^{--}$ given by (48.c) are proportional, to lowest order, to off-diagonal elements of the transfer matrix (32) in I,
which are linear in the site energies.  Also, as indicated earlier the form (32) of I of this matrix remains valid if,
even in the considered case of complex wavenumbers and corresponding complex tight-binding parameters in
(\ref{eq24}-\ref{eq25}), the symbols $O^*$, with $O\equiv s_j, u_j, v_{ij}, \omega_{ij}$, are taken to mean replacement
of the exponent coefficients $i=\sqrt{-1}$ in (31) of I by $-i$.  For these reasons we refrain from further
explicitating the reflection coefficients and the interchannel transmission coefficients in
(\ref{Appendix4}-\ref{Appendix5}).

We now turn to the discussion of the explicit forms of the intrachannel transmission coefficients,
$|t^{--}_{jj}|^2,j=1,2,3$, which play an important role and whose evaluation to second order requires more effort. 
Using the definition (48.e) and the explicit expressions of transfer matrix elements in (32) of I we obtain successively

\begin{align}\label{Appendix6}
\beta_1 &=
-e^{-i(k_2+k_3)N_L}\left[(1-i\sum_mb_{2m})(1-i\sum_n a_{3n})+\sum_{m,n}d_mq_ne^{i(m-n)(k_3-k_2)}\right]\;,\nonumber\\
\beta_5 &=
e^{-i(k_1+k_3)N_L}\left[(1-i\sum_ma_{1m})(1-i\sum_n a_{3n})+\sum_{m,n}p_mg_ne^{i(m-n)(k_1-k_3)}\right]\;,\nonumber\\
\beta_9 &=
-e^{-i(k_1+k_2)N_L}\left[(1-i\sum_ma_{1m})(1-i\sum_m b_{2m})+\sum_{m,n}c_mf_ne^{i(m-n)(k_2-k_1)}\right]\quad .
\end{align}

Similarly we evaluate the cubic form $\Delta$ defined in (48.f) and (45) of I to second order in the tight-binding
parameters in (32) of I.  This yields

\begin{multline}\label{Appendix7}
\Delta =
 -e^{-i(k_1+k_2+k_3)N_L}
\left[1-i\sum_m
(a_{1m}+b_{2m}+a_{3m})-\sum_{m,n}\right.\\
\left.
\biggl(a_{1m}b_{2n}+a_{3m}b_{2n}+a_{1m}a_{3n}
+g_mp_ne^{i(m-n)(k_3-k_1)}
+d_mq_ne^{i(m-n)(k_3-k_2)}+c_mf_ne^{i(m-n)(k_2-k_1)}\biggr)\right]
\quad .
\end{multline}

The atomic tight-binding parameters given in (24-25) of I are:

\begin{align}\label{Appendix8}
a_{1n}
&=
\frac{\varepsilon_{1n}+2\varepsilon_{2n}+\varepsilon_{3n}}{ 8\sin k_1}\;,\;
a_{3n}=\frac{\varepsilon_{1n}+2\varepsilon_{2n}+\varepsilon_{3n}}{ 8\sin
k_3}\quad ,\nonumber\\
b_{2n}
&=
\frac{\varepsilon_{1n}+\varepsilon_{3n}}{ 4\sin k_2}\;,\;
c_n=f_n=\frac{\sqrt 2(\varepsilon_{1n}-\varepsilon_{3n})}{ 8\sqrt{\sin
k_1\sin k_2}}\quad ,\nonumber \\
d_n=q_n
&=
\frac{\sqrt 2(\varepsilon_{1n}-\varepsilon_{3n})}{ 8\sqrt{\sin k_2\sin
k_3}}\;,\;
g_n=p_n=\frac{\varepsilon_{1n}-2\varepsilon_{2n}+\varepsilon_{3n}}{
8\sqrt{\sin
k_1\sin k_3}}\quad ,
\end{align}
for free boundary conditions, with the wavenumbers, real or imaginary, defined by (\ref{eq10}) and

\begin{align}\label{Appendix9}
a_{1n}
&=
\frac{\varepsilon_{1n}+\varepsilon_{2n}+\varepsilon_{3n}}{ 6\sin k_1}\;,\;
a_{3n}=\frac{2\varepsilon_{2n}+\varepsilon_{3n}}{ 6\sin k_2}\quad
,\nonumber\\
b_{2n}
&=
\frac{2\varepsilon_{1n}+\varepsilon_{3n}}{ 6\sin k_2}\;,\;
c_n=\frac{\varepsilon_{1n}-\varepsilon_{3n}}{ 6\sqrt{\sin k_1\sin k_2}}\;,\;
g_n=\frac{\varepsilon_{2n}-\varepsilon_{3n}}{ 6\sqrt{\sin k_1\sin k_2}}\quad
,\nonumber \\
d_n
&=
\frac{\varepsilon_{3n}-\varepsilon_{2n}}{ 6\sin k_2}\;,\;
f_n=\frac{2\varepsilon_{1n}-\varepsilon_{2n}-\varepsilon_{3n}}{ 6\sqrt{\sin
k_1\sin k_2}}\quad ,\nonumber \\
p_n
&=
\frac{-\varepsilon_{1n}+2\varepsilon_{2n}-\varepsilon_{3n}}{ 6\sqrt{\sin
k_1\sin k_2}}\;,\;
q_n=\frac{\varepsilon_{3n}-\varepsilon_{1n}}{ 6\sin k_2}\quad ,
\end{align}
for periodic boundary conditions where the wavenumbers are given by (\ref{eq11}).  Finally we insert (\ref{Appendix6})
and (\ref{Appendix7}) in (\ref{Appendix4}) for the intrachannel transmission coefficients $|t^{--}_{jj}|^2$ and
evaluate the resulting expressions to second order in the site energies.  This yields the relatively simple final
expressions

\begin{multline}\label{Appendix10}
|t^{--}_{11}|^2|e^{-ik_1N_L}|^2=1-2\;\text{Im}\;\sum_m
a_{1m}+\sum_{m,n}\left[a_{1m}a^*_{1n}-2\;\text{Re}\;(a_{1m}a_{1n})\right]\\
-\sum_{m,n}\left[\left(g_mp_ne^{i(m-n)(k_3-k_1)}+c_mf_ne^{i(m-n)(k_2-k_1)}\right)+c.c.\right]
\quad ,
\end{multline}

\begin{multline}\label{Appendix11}
|t^{--}_{22}|^2|e^{-ik_2N_L}|^2=1-2\;\text{Im}\;\sum_m
b_{2m}+\sum_{m,n}\left[b_{2m}b^*_{2n}-2\;\text{Re}\;(b_{2m}b_{2n})\right]\\
-\sum_{m,n}\left[\left(d_mq_ne^{i(m-n)(k_3-k_2)}+c_mf_ne^{i(m-n)(k_2-k_1)}\right)+c.c.\right]\quad ,
\end{multline}

\begin{multline}\label{Appendix12}
|t^{--}_{33}|^2|e^{-ik_3N_L}|^2=1-2\;\text{Im}\;\sum_m
a_{13m}+\sum_{m,n}\left[a_{3m}a^*_{3n}-2\;\text{Re}\;(a_{3m}a_{3n})\right]\\
-\sum_{m,n}\left[\left(g_mp_ne^{i(m-n)(k_3-k_1)}+d_mq_ne^{i(m-n)(k_3-k_2)}\right)+c.c.\right]\quad ,
\end{multline}
where summations over $m$ and $n$ run independently from $m=1$ to $m=N_L$ and from $n=1$ to $n=N_L$.


\begin{thebibliography}{unsrt}
\bibitem{1} J. Heinrichs, Phys. Rev. B{\bf 66}, 155434-1 (2002).
\bibitem{2} For an incisive recent discussion of the physics of the Landauer formula, see Y. Imry and R. Landauer, Rev.
Mod. Phys. {\bf 71}, S306 (1999).
\bibitem{3} See e.g. J.H. Davies, The Physics of Low-Dimensional Semiconductors (Cambridge University Press, Cambridge
1998).
\bibitem{4} R. Johnston and H. Kunz, J. Phys. {\bf 16}, 3895 (1983).
\bibitem{5} P.F. Bagwell, Phys. Rev. B{\bf 41}, 10354 (1990). 
\bibitem{6} M. Cahay, S. Bandyopadhyay, M.A. Osman and H.L. Grubin, Surf. Science {\bf 228}, 301 (1990).
\bibitem{7} S. Datta, M. Cahay and M. Mc Lennan, Phys. Rev. B{\bf 36}, 5655 (1987).
\bibitem{8} J.D. Jackson, Classical Electrodynamics, second edition (Wiley, New York, 1975).
\bibitem{9} Y. Imry, Introduction to Mesoscopic Physics (Oxford University Press, 1997).
\bibitem{10} M. B\"{u}ttiker, Y. Imry, R. Landauer and S. Pinhas, Phys. Rev. B{\bf 31}, 6207 (1985).
\bibitem{11} For the present case where wavenumbers may be pure imaginary the starred parameters in (30) and (32) of I
are to be regarded as meaning replacement of only those symbols $i=\sqrt{-1}$ appearing explicitly in the definitions
(\ref{eq31}) by $-i$.
\bibitem{12} N.M. Makarov and Yu. V. Tarasov, J. Phys.: Condens. Matter {\bf 10}, 1523 (1998); ibid. Phys. Rev. B {\bf 64},
235306-1 (2001).
\bibitem{13} C.W.J. Beenakker, Rev. Mod. Phys. {\bf 69}, 731 (1997).
\bibitem{14} M. Janssen, Phys. Rep. {\bf 295}, 1 (1998).
\bibitem{15} D.J. Thouless, J. Phys. C: Solid State Phys. {\bf 6}, 249 (1973).
\bibitem{16} D.J. Thouless, Phys. Rev. Letters {\bf 39}, 1167 (1977).
\end{thebibliography}
\end{document}